\documentclass{revtex4}

\usepackage{graphicx}
\def\beq{\begin{equation}}
\def\eeq{\end{equation}}
\def\calo{{\cal O}}
\newcommand{\roughly}[1]{\raise.3ex\hbox{$#1$\kern-.75em
\lower1ex\hbox{$\sim$}}}

\begin{document}
% You should use BibTeX and revtex.bst for references
\bibliographystyle{revtex}

% number  and your conference paper identification number on the
% title page in preprint mode. Multiple \preprint commands are allowed.
\preprint{hep-th/0110127}

\title{Black hole production in TeV-scale gravity, and the future of high
energy physics}
% Optional argument for running titles on pages
%\title[]{}

\author{Steven B. Giddings}
\email[]{giddings@physics.ucsb.edu}
%\thanks{}
\affiliation{Department of Physics\\ University of California\\
Santa Barbara, CA 93106}


\date{\today}

\begin{abstract}
If the Planck scale is near a TeV, black hole production should be possible
at colliders, as well as by cosmic rays.  I begin with a review of the
two approaches to TeV-scale gravity, large extra dimensions and warped
compactification, presented in a unified framework.
Then properties of such black holes and estimates of their
production rates are given, and consequences for the future of high-energy
experimental physics are discussed.  
\end{abstract}
% \pacs{}

%\maketitle must follow title, authors, abstract and \pacs
\maketitle

%\section{Introduction}
%\label{}
%\subsection{}
%\subsubsection{}

\section{Introduction}

The two most important -- and mysterious -- scales in physics are the
Planck scale, $G_N^{-1/2} \simeq 10^{19}$ GeV, and the weak scale,
$G_F^{-1/2}\simeq 300$ GeV.  The hierarchy problem is the problem of
explaining the large disparity between these scales.  In traditional
scenarios, the Planck scale is fundamental, and the weak scale is derived
from it via some dynamical mechanism.  However, we have recently begun
exploring an alternative viewpoint:  {\it the weak scale is the fundamental
scale of nature, and the four-dimensional Planck scale is to be derived
from that}.  Ingredients in constructing such a scenario include large or
warped extra dimensions, propagation of matter and gauge degrees of freedom
on brane worlds, and a {\it fundamental} Planck scale of $\calo(TeV)$.  I
will use the generic name ``TeV-scale gravity'' (TeVG) for these scenarios.

If the fundamental Planck scale is ${\cal O}(TeV)$, we are at the
threshold of a phenomenally exciting period in experimental physics.  In
particular, we might hope to observe experimentally strings, branes, 
Kaluza-Klein modes from the extra dimensions, and other quantum gravity 
phenomena.  But most remarkably -- and
largely independent of one's beliefs about the ultimate nature of quantum
gravity -- we would be able to produce black holes\cite{Giddings:2000ay,
Giddings:2001bu,Dimopoulos:2001hw}.  Their
production should be copious and would have outstanding signatures.  It
also appears to signal the end of our long quest to understand physics at
shorter distances.

In these proceedings I'll review some of these developments.  I'll begin
with a rapid review of TeV-scale gravity scenarios and their associated
theoretical challenges.  This review will present the two scenarios
for TeVG -- large extra dimensions and warped compactification -- in
a unified framework perhaps not widely appreciated. 
I'll then discuss black holes in these scenarios,
and their production in accelerators.  High energy cosmic rays also may
have sufficient energy to produce black holes, and I'll next summarize the
corresponding expectations.  This is followed by a discussion of the
consequences of black hole production for the future of high energy
physics.  Lastly, following the HEPAP charge, I'll briefly
address the implications of these possibilities for our future strategy in
experimental physics.  I've tried to include references necessary for
clarity; more complete references can be found in \cite{Giddings:2001bu}.

\section{TeV scale gravity}

Conventional compactification scenarios are now widely familiar.  We
imagine that in addition to the four spacetime dimensions we see, with
coordinates $x^\mu$, there are $D-4$ unseen dimensions with
coordinates $y^m$.  The full $D$-dimensional metric takes the form
\begin{equation}
ds^2=dx^\mu dx_\mu + g_{mn}(y) dy^m dy^n \ ,
\end{equation}
where the characteristic size of the extra dimensions is ${\cal O}(l_{Planck})$,
explaining their invisibility.

There is an important generalization of this that respects the
(approximate) 4d Poincar\'e invariance that we observe in nature: the scale
of the four-dimensional metric may vary depending on location in the extra
dimension,
\begin{equation}
ds^2=e^{2A(y)}dx^\mu dx_\mu + g_{mn}(y) dy^m dy^n \ ,\label{p3-28warpmet}
\end{equation} 
for some function $A(y)$.  Such a metric is known as a {\it warped metric},
and the factor $\exp\left\{2A\right\}$, which can alternately be thought of
as giving a position-dependent red shift, is known as a {\it warp factor}.

Given such a warped compactification, we would like to understand the
observed strength of 4d gravity.  Suppose that fundamental physics has an
effective action
\begin{equation}
S={1\over 8\pi G_D} \int d^D x \sqrt{-g} ~{1 \over 2} {\cal R} + \int 
d^D x \sqrt{-g} {\cal L}\ 
\label{p3-28Eact}
\end{equation}
where $G_D$ is the gravitational constant, ${\cal R}$ is the curvature
scalar, 
and ${\cal L}$ is the lagrangian
for non-gravitational fields.  For the Planck mass we will use a
convention useful in comparing to 
experimental bounds\cite{Peskin:2000ti} (for further
discussion comparing conventions see the appendix):
\beq
M_p^{D-2}={(2 \pi)^{D-4} \over 4 \pi G_D }\ .
\label{p3-28Mpnorm}
\eeq
If the metric (\ref{p3-28warpmet}) 
satisfies the equations of motion derived from
(\ref{p3-28Eact}), then four-dimensional 
metric fluctuations, with metric of the form
\beq
ds^2=e^{2A(y)}g_{\mu\nu}(x) dx^\mu dx^\nu + g_{mn}(y) dy^m dy^n \ ,
\eeq
are governed by the 4d action
\beq
S_4= {M_4^2 \over 4} \int d^4 x \sqrt{-g_4(x)} {\cal R}_4
\eeq
with the four- and D-dimensional Planck masses related by
\beq
{M_4^2\over M_p^2} = M_p^{D-4} \int {d^{D-4}y \over (2\pi)^{D-4}} \sqrt{g_{D-4}}
e^{2A}\equiv M_p^{D-4} V_w\  .\label{p3-28mplanck}
\eeq
This equation defines the ``warped volume'' $V_w$.

Eq.~(\ref{p3-28mplanck}) is key to understanding the relationship between 
different scenarios.  There are two obvious possibilities:

\begin{enumerate}

\item Conventional small-scale compactification:  
$M_p\sim M_4 \sim 10^{19} GeV$, and $V_w\sim 1/M_p^{D-4}$.
\item TeV-scale gravity scenario:  $M_p\sim 1$ TeV, which then requires
$V_w\gg 1/M_p^{D-4}$.
\end{enumerate}

There are two basic kinds of schemes to achieve a large warped volume.  
The first is the scenario of Arkani-Hamed, Dimopoulos, and 
Dvali\cite{Arkani-Hamed:1998rs}, which imagines negligible warping,
$e^A\approx 1$, and simply large volume, $M_p^{D-4} V_{D-4}\gg1$.  The size
of the extra dimensions then ranges from $\calo(mm)$ for $D=6$ to
$\calo(10\,fm)$ for $D=10$.  However, gauge interactions have been
well tested, with no evidence of extra dimensions, to around 100 GeV.
These statements can be reconciled by noting that string theory provides a
natural mechanism for gauge interactions to operate in a lower-dimensional
arena: they can propagate on a D-brane.  So the picture is that of large
extra dimensions in which only gravity propagates, and a typically
three-dimensional D-brane on which fermions and gauge bosons propagate.
%\begin{figure}
%\includegraphics{add.eps}
%\caption{In the ADD scheme, gravity propagates in large flat extra
%dimensions, while fermions and gauge bosons propagate on a brane.}
%\label{addpic}
%\end{figure}

The second scheme is based on large warp factor, with gauge and matter
fields propagating on branes as in the ADD scenario\cite{Arkani-Hamed:1998rs}. 
A general class of string
solutions
with requisite warp factor have been found in \cite{Giddings:2001yu} and
were discussed in Kachru's talk; toy
models with some of the basic features appear in the work of Randall
and Sundrum\cite{Randall:1999ee}.  See fig.~1.
\begin{figure}
\includegraphics{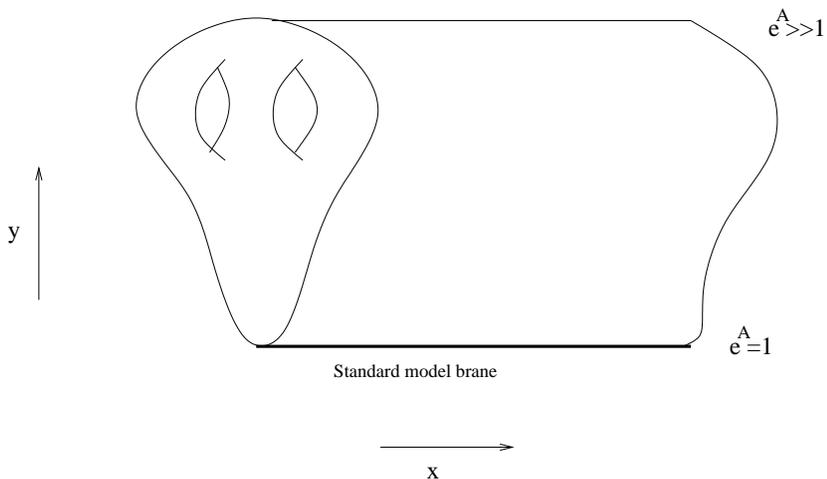}
\caption{Schematic of a warped compactification.}
\label{p3-28wcpic}
\end{figure}

Specifically, the gravitational action in (\ref{p3-28Eact}) has a 
``symmetry'' corresponding
to a global choice of scale:
\beq
g\rightarrow \lambda^2 g\ ;\ M_p\rightarrow M_p/\lambda
\eeq
where $g$ is the full $D$-dimensional metric, and with corresponding
scalings of the matter fields and other dimensionful parameters.  Since
$g_{\mu\nu} = e^{2A} \eta_{\mu\nu}$, 
this may be used to set $e^A=1$ at the brane where the standard model
propagates.  A large $e^A$ elsewhere in the compact dimensions can then 
yield a
large warped volume.  This is precisely what happens in  
\cite{Giddings:2001yu}.  These models also break supersymmetry, and have
vanishing cosmological constant at tree-level.  They  differ from the ADD 
models\cite{Arkani-Hamed:1998rs} in several important respects\cite{DeGi}, most
notably their Kaluza-Klein spectrum.

Clearly a critical question regards the likelihood that TeV-scale gravity
is realized, in one of the above scenarios or in a completely different
manner, in nature.  The scenarios based on brane worlds and large/warped
extra dimensions do face several challenges:
\begin{enumerate}
\item How does one obtain the standard model gauge group, $SU(3)\times
SU(2)\times U(1)$, and the correct matter representations?
\item Why does such a model reproduce the relationship between the coupling
constants that in traditional SUSY/GUT 
scenarios emerges from coupling unification?
\item Why is the proton stable?
\item How is the correct scale produced for neutrino masses?
\item What stabilizes moduli, such as the size of the extra dimensions?
\item What is the role of supersymmetry -- is it for example in protecting
the largeness of the extra-dimensions or warp factor?  How is it broken?
\item How do we obtain a small cosmological constant?
\end{enumerate}

Several of these are equally problematical for conventional SUSY/GUT
scenarios, {\it e.g.} based on small-scale string compactification.
However, so far TeVG  scenarios face additional theoretical
challenges such as 1)-4), and to some degree 6).  There are, however,
ideas that begin to address these, and moreover the space of
of such models has been less explored than SUSY/GUT scenarios.  
It may be that further
exploration reveals more natural solutions to these problems.  

\section{Black holes in brane-world scenarios}

\begin{figure}
\includegraphics{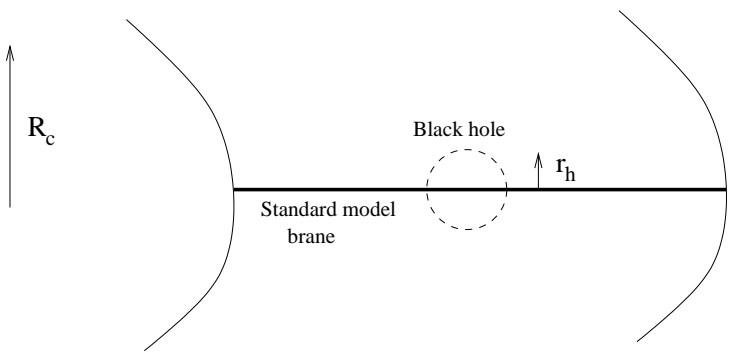}%
\caption{Shown is a black hole on a brane in a compactification with large or
warped extra dimensions.  We consider the approximation where the black
hole size is small as compared to characteristic geometrical scales.}
\label{p3-28bhfig}
\end{figure}

If TeV-scale gravity is realized in nature, production of black holes
should be possible for $\sqrt{s}\gg 1$ TeV.  Let's briefly consider
their properties.  We use two approximations.  The first is that we
initially 
assume
that black hole radii are small as compared to the radii and curvature
radii of the extra dimensions, and the scale on which the warp factor
varies:
\beq
r_h\ll R_c\ ,
\eeq
where $R_c$ denotes a characteristic geometrical scale; see fig.~2.
Secondly, the standard model brane has a tension and thus a gravitational
field.  However, we will consider black holes with masses typically larger
than the tension, and so neglect the effects of this gravitational
field\footnote{The authors of \cite{Arkani-Hamed:1998rs} neglect this
effect entirely.}.
These
approximations mean we effectively 
consider black holes in $D$-dimensional flat
spacetime.  Furthermore, as we'll discuss, we are interested in spinning
solutions.  These higher dimensional spinning black hole solutions were given by
Myers and Perry in \cite{Myers:1986un}, and are parametrized by their
mass $M$ and spin $J$.  Other parameters are given in terms of these in 
\cite{Myers:1986un}.
%, and require the formula for the volume of the unit
%$D-2$ sphere:
%\beq
%\Omega_{D-2}= { 2\pi^{(D-1)/2} \over  \Gamma({D-1 \over 2})}\ .
%\eeq
In the $J=0$ limit they take the form 
\beq
r_h(M,J) \,\, \buildrel J \rightarrow 0 \over\longrightarrow \,\,
2 \left[{ \pi^{D-7\over 2} \Gamma\left({D-1\over 2}\right)\over (D-2)} {M \over
M_p^{D-2}}\right]^{1/(D-3)}\ ,
\eeq
for the radius,
\begin{equation}
T_H 
\buildrel J\rightarrow 0 \over\longrightarrow \,\,
{D-3\over 4\pi r_h}\ 
\label{p3-28Thawk}
\end{equation}
for the Hawking temperature,
and
\begin{equation}
S_{BH} = 
\buildrel J\rightarrow 0 \over\longrightarrow\,\,
{\pi^{(D-1)/2}  \over 2\Gamma\left({D-1\over 2}\right) }{r_h^{D-2}\over G_D}\ 
\end{equation} 
for the entropy.

A critical question is at what mass is the black hole description valid.
In our conventions, the experimental bounds on the Planck mass in ADD
scenarios from absence of missing
energy signatures\cite{Giudice:1998ck,Mirabelli:1998rt,Peskin:2000ti}
are $M_p>1.1$ TeV -- .8 TeV for
$D= 6\,-10$.  Similar bounds appear in the RS toy models for warped
compactifications\cite{Davoudiasl:1999jd}.  In order
to neglect quantum or classical ({\it e.g.} string) effects that strongly
modify solutions, we must consider black holes at higher masses.  The
dominant quantum effect is the Hawking radiation, and criteria for this to
be small include a lifetime long as compared to $M^{-1}$, or validity of
the statistical description for the black hole.  The latter is more
stringent, and since $S_{BH}$ parameterizes the number of degrees of freedom
of the black hole, becomes
\beq
1\gg {1\over \sqrt{S_{BH}} }\ .
\eeq
For $M=5M_p$, $1/\sqrt{S_{BH}}\sim 1/5$, and for $M=10M_p$, 
$1/\sqrt{S_{BH}}\sim 1/8$, so we only trust a black hole description
beginning in this range.

Classical modifications to gravity, such as string theory, can also be
important.  String effects are relevant for $M\sim g_s^{-7/4}M_p$, where $g_s$
is the string coupling, and thus depend on the value of the coupling.
Weaker couplings imply higher thresholds for black hole behavior.
However, the string correspondence principle\cite{Horowitz:1997nw} states
that the black hole spectrum matches onto the string spectrum for $r_h$ of
order the string scale, suggesting that a black hole description is
essentially valid until this point. 

\section{Black hole production at accelerators}

The energy frontier is at hadron machines, and in hadron scattering
the black hole cross section is found from the partonic
cross section for partons $i$ and $j$ to form a black hole:
\begin{equation}
\sigma_{pp \rightarrow bh}(M_{min},s) =
  \sum_{ij} \int_{\tau_m}^1 d \tau \int_{\tau}^1 {dx \over x}
    f_i(x) f_j(\tau/x)
         \sigma_{ij \rightarrow bh}(\tau s)\ . \label{p3-28ppcross}
\end{equation}
Here $\sqrt s$ is the collider center of mass energy, $x$ and $\tau/x$
are 
the parton
momentum fractions, and $f_i$ are the parton distribution functions.  
Studies should include the parameter $M_{min}$
corresponding to the minimum mass for a valid black hole description, and
we define $\tau_m = M_{min}^2/s$.

To estimate the parton-level cross section, consider partons scattering at
center of mass energy $\sqrt{s}$ and impact parameter $b$.  A longstanding
conjecture in relativity is Thorne's hoop conjecture\cite{Thorne:1972ji}, which
states that horizons form when and only when a mass $M$ is compacted into a
region whose circumference in every direction is less than $2\pi r_h(M)$.
This conjecture implies that the cross-section for black hole production is
\beq
\sigma_{ij \rightarrow bh}(s)\sim \pi r_h^2(\sqrt{s})\ . \label{p3-28partcross}
\eeq

Can we believe this conjecture?  High-energy gravitational scattering has
been studied at the classical level in four dimensions; 
in the symmetrical case of zero impact
parameter, Penrose\cite{Penrose} has found an apparent horizon of area ${\cal
A} = 8\pi s$, implying that a black hole of mass $M>\sqrt{s/2}$ forms.
d'Eath and Payne\cite{D'Eath:1992hb,D'Eath:1992hd,D'Eath:1992qu} extended
these results, arguing using a perturbative analysis that a black hole of
mass $M\sim .84 \sqrt{s}$ forms.  The fact that big black holes form in
on-axis collisions strongly indicates that black holes should form for
impact parameters $b\roughly<r_h$, but clearly further study, which may
have to be numerical, is desired.

The question of quantum corrections is also addressed by the presence of a
large horizon.  This in particular suggests that the curvature at the
horizon is small.  In the absence of strong curvature, a semiclassical
treatment of gravity should be valid.  The semiclassical approximation only
appears to fail near the center of the black hole, long after a black hole
has formed.  These points can be illustrated in four dimensions 
by imagining a collision of
partons with CM energy equal to the rest mass of the sun; here we'd expect
a  horizon  of radius $\sim 1$ km 
to form, with weak curvature at the horizon.
This buttresses the argument for (\ref{p3-28partcross}).

The estimate (\ref{p3-28partcross})   has 
been criticized by Voloshin in \cite{Voloshin:2001vs}, where two
objections are raised.
The first is a suggestion that the calculation of
the rate should include the exponential of minus the euclidean 
black hole action.
However, this seems clearly incorrect:  as described above, black hole
formation is a {\it classically allowed} (in fact compulsory!) process.  One
ordinarily encounters the euclidean action only when studying amplitudes
for quantum processes that are classically forbidden.  The second objection
involves an application of CPT to argue that since the decay $BH\rightarrow
ij$ is small (thermally suppressed), 
the amplitude $ij\rightarrow BH$ should be small.  This however neglects
the fact that the black hole should have a number of distinct states.  We
can label these as $\alpha$, and the statement that $\alpha\rightarrow ij$
is small CPT conjugates to the statement that ${\bar{i} }{\bar{j} }
\rightarrow {\bar \alpha}$ is small.  This does not mean that
$ij\rightarrow \alpha$ is small; in particular, in classical gravity, the
time reverse of a black hole is a white hole which is a very different
state.  Application of such state-counting arguments clearly requires an
improved
understanding of the description of the internal states
of a black hole.

Another important point is that typically black holes are produced with
large spin.  This is because the cross-section is dominated by large impact
parameters, so typically $J\sim r_h M$.  This complicates study of the
formation process, and in particular indications from other examples of
gravitational collapse with high angular momentum suggest the possibility
of added complexities such as initially toroidal horizons\footnote{I thank
Scott Hughes for conversations on this point.}.
The differential cross-section can
be parametrized as 
\beq
{d\sigma_{ij}(s)\over dJdM}= F(s,J,M) \pi r_h^2\ ,
\eeq
where further effort is required to compute the function $F$.  However, we
believe that the total cross-section can be approximated by 
(\ref{p3-28partcross}).

We therefore estimate rates using (\ref{p3-28partcross}) and the CTEQ5
structure functions\cite{Lai:1999wy}\footnote{We thank T. Rizzo for performing
these estimates.}.  The result is impressive.  For example, assume a Planck
mass
$M_p= 1$ TeV, and consider LHC with $\sqrt{s}=14$ TeV.    If the minimum
mass to produce true black holes is $M_{min}=5$ TeV, they are produced with
cross-section $\sim 2.4 \times 10^{5}$ fb and thus at a rate $\sim 1$ Hz.
If 10 TeV is required to make a black hole, the cross section and rate
are still respectable at $~10$ fb and 3/day.  Looking further into the
possible future, if VLHC were built at $\sqrt s = 100 $ TeV and luminosity
$100 fb^{-1}/yr$, 10 TeV black holes would be produced at $\sim kHz$ rates,
and 50 TeV black holes at $\sim .5$ Hz.
Notice that the cross section grows
as 
\beq
\sigma\sim s^{1\over D-3}\label{p3-28crossg} \ .
\eeq
Thus at sufficiently high energies
colliders inevitably become black hole factories, and ultimately black hole production
becomes a dominant process.  The critical question, depending on the value
of the Planck scale, is what energy reach we require to begin to see these
stunning developments.  It is amusing to note that if we're lucky, colliders
could beat LIGO to observation of black hole formation.

\section{Black hole decays and signatures}

Once produced, black holes decay primarily via Hawking radiation, and
should yield events that stand out in detectors.  These decays and their
signatures were surveyed in Scott Thomas' talk, and are discussed in detail
in \cite{Giddings:2001bu}, but I'll briefly summarize some of the most
notable points here.

Black hole decay occurs in several stages, with different characteristic
time-scales and energy spectra.  When a black hole first forms in a
high-energy collision, the horizon will be highly asymmetrical, and could
even be topologically non-trivial.  The black hole will then settle down to
a symmetrical rotating black hole by emitting gauge and gravitational
radiation.  In the course of this emission, the horizon can only grow, by
the area theorem.  Since the final state of this phase is a black hole with
no hair, we refer to this as the {\it balding} phase.  The duration of this
phase is expected to be $\calo(r_h)$, and the characteristic frequency of
the radiation emitted should be $\calo(1/r_h)$.  Based on the estimates of
\cite{D'Eath:1992hb,D'Eath:1992hd,D'Eath:1992qu},
in a head on collision one expects about $16\%$ of CM energy of the
partons to be emitted this way, with the likelihood of greater emission from
balding at larger impact parameters.

The next phase is {\it spin down}.  The black hole Hawking radiates, first
shedding its angular momentum by preferentially emitting quanta with
angular momenta $l\sim1$.  These quanta will have characteristic energies
given by the Hawking temperature $T_H$ at the time they are emitted.  This
process has been treated in detail for four-dimensional black holes by Page
\cite{Page:1976df,Page:1976ki}.  
Rough estimates based on this suggest that $25\%$ of the
black hole's energy is radiated in this phase, although Page's calculations
should be redone in the higher-dimensional context.

Spin down leaves behind a Schwarzschild black hole, which then continues to
Hawking radiate through what we call the {\it Schwarzschild phase}.  Here
again at a given instant quanta are emitted with a thermal spectrum at $\sim T_H$.  
As the black hole decays its temperature increases according to 
(\ref{p3-28Thawk}).  The
total spectrum of the decay products can be obtained by integrating the
thermal spectrum over this evolution.

Once the black hole reaches a mass $M\sim M_p$, Hawking's calculations
fail.  We call this phase the {\it Planck phase}.  One expects the final
decay of the Planck phase to result in emission of a few quanta with
energies $\calo(M_p)$.

These decays should be quite spectacular.  In particular, black hole events
should produce a large number, of order $S_{BH}\roughly>25$, of hard quanta,
with energies approaching a sizeable fraction of 1 TeV.  In particular, a
substantial fraction of the beam energy is thereby deposited in visible
transverse energy, in an event with high sphericity.  
Based on \cite{Page:1976df,Page:1976ki}, one can estimate that the ratio
of hadronic to leptonic activity is around 5:1.  (A more careful estimate
based on a higher-dimensional version of the analysis of 
\cite{Page:1976df,Page:1976ki} should
be done.)  Furthermore, it should ultimately be possible to determine the
black hole spin axis by observing the characteristic dipole pattern from
the spin-down phase.  More detailed discussion of black hole signatures can
be found in \cite{Giddings:2001bu}.

\section{Black holes from cosmic rays}

Physicists be warned: journalists regularly read our electronic archives!
After \cite{Giddings:2001bu} appeared, a
journalist almost immediately asked me the question, what if Hawking's
calculations are wrong, and black holes don't evaporate?  Of course, we
certainly believe that Hawking's calculations are correct, if not to the
last detail, and furthermore on general quantum grounds black holes should
decay -- they are massive states with no conserved quantities to stabilize
them.  But further assurances are welcome.

For energies accessible in the foreseeable future, an answer comes from cosmic
rays, which are observed up to lab energies $~10^{11}$ GeV.  They collide
with protons in the atmosphere, and therefore probe CM energies up to
$\sqrt{s}\sim 400$ TeV.  So if accelerators can investigate black hole
production, black holes are already being produced in the atmosphere; if
this weren't a safe thing to do, we wouldn't be here to talk about it.

We might hope to observe these events at cosmic-ray observatories, and thus
need rates.
At ultra-high energies, it is uncertain what fraction of 
cosmic-ray primaries are
nucleons versus heavy nuclei such as iron, and other constituents have not been
conclusively ruled out.  The most optimistic case for
producing black holes is nucleons; otherwise the following estimates have
to be readjusted to account for the distribution of the energy between the
constituents of a nucleus.  (This reduces the effective flux at a given
$pp$ CM energy.)
Suppose for 
example $M_p=1$ TeV and the black hole threshold is 10 TeV, and consider
the cosmic ray flux at $\sqrt{s}\sim 40$ TeV (thus $E_{lab}\sim
10^{18}eV$).  The results of \cite{Giddings:2001bu} show that with these
parameters the branching ratio for $pp\rightarrow BH$ is $\sim 3\times
10^{-9}$, resulting in roughly 100 black holes produced over the surface of
the earth in a year.  The rates are too small to be observed because $pp$
collisions are dominated by QCD processes.

This suggests that we consider cosmic rays with small standard-model cross
sections, in particular 
neutrinos\cite{DGRT,Feng:2001ib,Emparan:2001kf}.  
The ultra-high-energy neutrino flux is
not known, but {\it if} the ultra-high energy cosmic ray primaries are
dominantly protons, a lower bound is believed to follow from the observation
that these protons should scatter off the microwave background,
resonantly producing pions and hence
neutrinos\cite{Greisen:1966jv,Stecker:1979ah,Hill:1985mk}.  This is the physics
behind the GZK bound.  There may be other fluxes in addition to these
Greisen neutrinos, due to active galactic nuclei or gamma ray bursts
(for summaries, see 
\cite{Capelle:1998zz,Cline:1999ez,Sigl:2001th}).

The cross section for $\nu p\rightarrow BH$ can also be estimated from 
(\ref{p3-28partcross})
and the structure functions from an 
expression analogous to (\ref{p3-28ppcross}).  This
results\cite{DGRT,Feng:2001ib,Emparan:2001kf} 
in cross-sections $\sigma\sim 10^{6}$ pb for $\sqrt{s}\sim 100$ TeV in 
the optimistic case of $M_p= 1$ TeV and minimum black hole mass 5 TeV.

The Greisen flux peaks at $\sqrt{s}\sim 100$ TeV, and combined with the
above cross section, yields an estimated production rate
\beq
R\sim { {\rm several\ black\ holes}\over ({\rm year})(km^3 (we))}\ 
\eeq
where $we$ denotes water equivalent.
This appears above the threshold of detectability 
by the Hires Fly's Eye experiment, with acceptance $~1 km^3(we)$,
or the Auger detector, presently under construction,
with acceptance $~1 km^3(we)$ for its ground array; for the latter, such
estimates give 6--17 events/yr for $D=$5--10\footnote{Note that
\cite{Feng:2001ib} sets the Planck mass to 1 TeV in the conventions of
\cite{Dimopoulos:2001hw}, which in $D=10$ corresponds to roughly three
times the experimental bounds\cite{Peskin:2000ti} -- 
see appendix.  Feng and Shapere also
consider minimum black hole masses as low as 1 TeV, which in our
conventions correspond to $M\sim M_p/3$ -- far out of the range $M\roughly>
5M_p$ where the
black hole approximation is valid.  The preceding improved estimate,
provided by J. Feng, was recomputed with $M_p=1$ TeV and $M_{min}=5$ TeV.}.
(Some discussion of
possible signatures of such events appears in \cite{Anchordoqui:2001ei}.)
Note, however, some caveats. First, there could well be a mild
numerical suppression in (\ref{p3-28partcross}); 
for example, a factor $\calo(1/10)$ is very
significant in the cosmic ray context, but not in that of collider production.
Second, the presence of a Greisen neutrino flux at this level relies on the
assumption that charged cosmic ray primaries are protons, not nuclei.

Other planned detectors may be able to improve on this.  The proposed
upgrade to AMANDA, Icecube, instruments $~1 km^3(we)$ but faces issues in
resolving ultra-high energy events; OWL/AirWatch has potential reach to
$6\times 10^4 km^3(we)$.  
And, of course, other components to the neutrino flux enhance the odds; for
example, some models\cite{Stecker:1996th} 
of active galactic nuclei predict fluxes $10^5$ times
the Greisen flux, peaking around $\sqrt{s}=10$ TeV.  

It is also worth point out that if the Planck scale is beyond the reach of
LHC or a linear collider, $M_p\roughly> 6$ TeV, black hole events might
nonetheless
be
observed in cosmic ray experiments with sufficient acceptance.

\section{Consequences for high energy physics}

We do not know the ultimate theory of quantum gravity, although a good
guess is string theory.  However, one thing seems clear:  once we reach the
threshold to produce black holes, it will be very difficult and likely
impossible to probe shorter distances via high energy scattering.

Of course, the black-hole threshold is above the Planck mass, and it's
widely believed that sub-planckian distances can't make sense in quantum
gravity.  But suppose that nevertheless shorter-distance physics did
exist.  Black hole production would then render it invisible.  This is
because we need to perform scattering at energies $\gg M_p$ in order to see
such physics.  However, at these energies, a large black hole will form,
and cloak any hard process behind the horizon.  All we see is that a black
hole forms, and then evaporates via Hawking decay.  For larger
energies, we just get larger black holes.  This is directly related to
ideas about the infrared/ultraviolet connection that have been widely
discussed in the theoretical literature. 

Black hole production therefore represents the end of short distance
physics.  Fortunately, it is not the end of high energy physics.  As we go
to higher energies, the black holes that we make get larger and extend
further into the extra dimensions.  At some point
they get large enough to run into other features of the extra dimensions.
For example, they might encounter the finite radius of one of the
dimensions, or finite curvature radii, or bump into other branes in the
extra dimensions.  As the black holes become large enough to detect these
features, their cross-sections, decay rates, and 
decay spectra change.  For example, once
a black hole has a radius larger than that of one of the extra dimensions,
or larger than a curvature radius in the extra dimensions, the effective
dimension in the production cross-section (\ref{p3-28crossg}) 
changes.  By measuring
kinks in the cross-section at larger energies, one can explore the extra
dimensions.  So high energy experiments  will be used to study the geography of the extra
dimensions.  

\section{Strategies for the future}

If nature realizes TeV-scale gravity, we are at the threshold of a
phenomenally exciting period of physics.  We will finally be able to
experimentally address
the puzzles of quantum gravity, and we should start making black holes, close
the frontier of short-distance physics, and instead begin
exploration of 
the terra incognita of extra dimensions.  At the same time experimental
physics may reveal exciting discoveries such as strings, branes, or other
exotica of a fundamental theory of gravity. 
What are the odds that gravity is
realized this way, and what should we do to be sure we don't miss out on
such a discovery?

An amusing exercise has been polling some ($\sim 10$) 
of my fellow theorists to see
what odds they would assign to the various possibilities once we understand
the physics of the TeV scale.  They were given the following four choices,
and the range of odds given were:  TeV-scale gravity 0-25\%, SUSY
25-100\%, just the standard model 0-30\%, and none of the above (or other)
5-65\%.  

Present theoretical wisdom clearly holds that the most likely discoveries at the
TeV scale are the Higgs and supersymmetry.  The case for this is buttressed
by various indirect arguments, such as the apparent unification of
couplings.  If we review the theoretical issues for TeV-scale gravity
scenarios discussed in section II, with our present state of knowledge it
does appear more difficult for TeV-scale gravity to fit nature than
supersymmetric grand unification to do so -- though they both confront theoretical
obstacles.  

There are counterpoints to this.  First, TeV-scale gravity is much younger
and less explored -- it may become more attractive if we gain deeper
insight that leads to resolution of some of its difficulties.  Secondly,
let me define an index, along the lines of Gross' talk in the evening
sessions, that may serve as a guide towards the importance of investigating
a given scenario:
\beq
I=({\rm probability})\cdot ({\rm impact\ of\ discovery})\ .
\eeq
While the probabilities that people assign to TeV-scale gravity are
substantially lower than just supersymmetry (and of course they are not
mutually exclusive), the impact of discovering strong gravity at the TeV
scale would be far greater. 

What do we need to do to discover or rule out TeV-scale gravity?   
On the experimental side, LHC
should give a bound\cite{Peskin:2000ti,Abe:2001nq} 
$M_p\roughly> 6$ TeV, and a linear collider would not appear to reach
beyond\cite{Abe:2001nq}.  
On the
theoretical side, the possibility that $M_p>6$ TeV seems like a definite
possibility, but clearly more theoretical understanding of these scenarios
is needed.

So what is a reasonable experimental course of action?  Since supersymmetry
is the best bet and a linear collider will likely be crucial in exploring
it, building a linear collider seems like a good next step, and supplies an
alternative approach to placing bounds on TeVG\cite{Abe:2001nq}.
However, the much more spectacular scenario of
strong gravitational physics (or something more bizarre) 
lying somewhere not far above a TeV 
is
a definite possibility.  We should hedge our bets, certainly
by continued theoretical exploration of these scenarios.  But more
importantly, we should actively 
continue to pursue long range plans that will allow us
to push the energy frontier beyond that explored by LHC.

\begin{acknowledgments}
I'd like to thank L. Bildsten, J. Feng, 
J. Hewett, G. Horowitz, S. Hughes, and 
R. Myers
for very useful discussions.  Thanks also go to T. Rizzo for discussions
and cross-section computations, and especially to my collaborator
S. Thomas for many useful discussions.  
This work was supported in part by the DOE under contract 
DE-FG-03-91ER40618.
\end{acknowledgments}

\appendix
\section{Comparison of Conventions}

In this proceedings and \cite{Giddings:2001bu} we normalize $M_p$ in a
convention useful in quoting experimental bounds\cite{Peskin:2000ti}.  
In these conventions,  the $D$-dimensional Newton
constant and the Planck mass are related by
\beq
M_p^{D-2} = {(2\pi)^{D-4}\over 4 \pi G_D}\ .
\eeq
At least two other conventions exist.  For example, bounds quoted in the
{\it Linear Collider physics resource book}\cite{Abe:2001nq} are quoted for
$M_D$ in
the convention of \cite{Giudice:1998ck}:
\beq
M_D^{D-2}= {(2\pi)^{D-4}\over 8 \pi G_D}\ .
\eeq
Thus
\beq
M_p = 2^{1\over D-2} M_D\ ,
\eeq
a small relative correction.

The paper by Dimopoulos and Landsberg \cite{Dimopoulos:2001hw} uses
somewhat different conventions,
\beq
M_{DL}^{D-2}= {1\over G_D}\ .
\eeq
Therefore the relation between the Planck masses in our two normalizations
is
\beq
M_{p}^{D-2}= 2^{D-6} \pi^{D-5} M_{DL}^{D-2}\ .
\eeq
In $D=6$ the difference is not great, $M_{p}= 1.3 M_{DL}$, but in $D=10$
the difference results in a substantial factor: $M_{p}= 2.9 M_{DL}$.

% figures should be put into the text as floats.
% Use the graphicx package (distributed with LaTeX2e).
% See the LaTeX Graphics Companion by Michel Goosens, Sebastian Rahtz,
% \begin{figure}
% \caption{}
% \label{}
% \end{figure}

% tables follow here or maybe be put in the text
% \begin{table}
% \caption{}
% \label{}
% \end{tabular}
% \end{table}

\bibliography{p3-28}

\end{document}